\documentstyle[12pt,aasms4]{article}

\def\etal{{\it et al.}}

\def\eg{{\it e.g.}}
\def\lap{\hbox{${_{\displaystyle<}\atop^{\displaystyle\sim}}$}}
\def\gap{\hbox{${_{\displaystyle>}\atop^{\displaystyle\sim}}$}}

\begin{document}

\title{Are We Seeing Magnetic Axis Reorientation in the Crab and 
Vela Pulsars?} 
\author{Bennett Link\altaffilmark{1}}
\affil{Montana State University, Department of Physics, Bozeman MT
59717; blink@dante.physics.montana.edu} 
\altaffiltext{1}{Also Los Alamos National Laboratory}
\and
\author{Richard I. Epstein}
\affil{Los Alamos National Laboratory, Mail Stop D436, Los Alamos, NM
87545; epstein@lanl.gov}

\begin{abstract}

Variation in the angle $\alpha$ between a pulsar's rotational and
magnetic axes would change the torque and spin-down rate. We show that
{\em sudden} increases in $\alpha$, coincident with glitches, could be
responsible for the persistent increases in spin-down rate that follow
glitches in the Crab pulsar. Moreover, changes in $\alpha$ at a rate
similar to that inferred for the Crab pulsar account naturally for the
very low braking index of the Vela pulsar. If $\alpha$ increases with
time, all pulsar ages obtained from the conventional braking model are
underestimates. Decoupling of the neutron star liquid interior from
the external torque cannot account for Vela's low braking index.
Variations in the Crab's pulse profile due to changes in $\alpha$
might be measurable. 

\end{abstract}


\section{Introduction}

Although the rotational behavior of pulsars has been monitored in
detail for decades, some aspects of how an isolated pulsar spins down
are unresolved. Questions surround the manner in which angular
momentum is removed from the system and how the different components
of the neutron star interior couple to one another.  In particular,
most pulsars do not slow in a regular fashion, but undergo variations
in their spin rates in the form of glitches and timing noise. Except
for such timing irregularities, a pulsar is expected to slow down
steadily. For example the {\em vacuum dipole model}, a theory of
rotational energy loss to magnetic dipole radiation
(\cite{dipole_model}), predicts $\dot{\Omega}\propto -\Omega^3$, where
$\Omega$ is the pulsar rotational velocity. This spin-down law gives a
braking index of $n_{\rm obs}\equiv
\Omega\ddot{\Omega}/\dot{\Omega}^2=3$. Timing irregularities make
meaningful determination of $\ddot{\Omega}$ and hence $n_{\rm obs}$
difficult or impossible in most cases. Consequently, the only braking
indices available until recently were those for three very young,
relatively quiet pulsars: the Crab pulsar ($n_{\rm obs}=2.51\pm 0.01$;
\cite{LPS}), PSR B1509-58 ($n_{\rm obs}=2.837\pm 0.001$;
\cite{psr1509_index}), and PSR B0540-69 ($n_{\rm obs}=2.24\pm 0.04$;
\cite{psr0540_index}). These braking indices are in significant
disagreement with the vacuum dipole model expectation. Plasma in the
pulsar magnetosphere could carry angular momentum from the star and
alter the magnetic structure from the dipolar configuration, giving a
braking index as low as one (see, \eg, \cite{wind}). Another
possibility is that the magnetic moment of the star changes in time,
through either a change in the surface field strength (see, \eg,
\cite{field_growth}; \cite{MP}) or the angle between the magnetic and
spin axes (see, \eg, \cite{counteralignment}; \cite{LEB}; \cite{PP}).

In the Crab pulsar, persistent increases in the spin-down rate are
observed to accompany glitches (see, \eg, \cite{LPS} and Table 1).  This
phenomenon can in principle be explained by either a sudden change in
the external torque or in the moment of inertia acted upon by the
torque. In this Letter we suggest that the persistent increases in the
Crab's spin-down rate reflect sudden, glitch-induced reorientations of
the star's magnetic axis that increase the external torque. 

Recently, Lyne \etal\ (1996) attempted to separate the glitch activity
of the Vela pulsar from its underlying spindown, and obtained a 
surprisingly small value of $\ddot{\Omega}$. If this small
$\ddot{\Omega}$ reflects the true spin-down of the star, the braking
index is only $1.4\pm 0.2$. This braking index, far smaller than those
for younger pulsars, suggests that the braking index decreases with
time.  This possibility forces reconsideration of traditional
determinations of pulsar ages that assume constancy of the pulsar
magnetic moment.  In this Letter we demonstrate that shifts in
$\alpha$ at a rate similar to that inferred for the Crab account
naturally for Vela's low braking index. 

\section{The Crab Pulsar}

The Crab pulsar has produced 6 glitches since 1969 (\cite{LPS}).
During each glitch, the spin rate suddenly increases by $\Delta\Omega$
(see column 2 of Table 1). Immediately following each glitch, the
magnitude of the spin-down rate $\dot{\Omega}$ is larger than
before. Most of the excess spin-down rate decays within weeks of the
glitch. However the long-term behavior, on which we focus here, shows
a persistent shift (or {\em offset}) $\Delta\dot{\Omega}$ in the
spin-down rate that does not decay (see column 3 of Table 1). This
type of behavior was most striking after the 1989 glitch, in which the
persistent shift in spin-down rate was especially large (see
Fig. 1). Following the glitch, the spin-down rate remained larger than
that expected from an extrapolation of the preglitch behavior. In all
the glitches, the shifts in the spin-down rate persisted to the next
glitch and were of the same sign.  The cumulative effect of these
offsets amounted to a total increase in the star's spin-down rate of
0.07\% over 23 years.

Any proposed mechanism for these persistent spin-down rate offsets
must conserve angular momentum.  The
evolution of the angular velocity $\Omega (t)$ of a rigidly rotating
pulsar under an external torque $N_{\rm ext}$, is 
\begin{equation}
{d(I\Omega)\over dt} = -N_{\rm ext}, 
\label{cons} 
\end{equation} 
where $I$ is the total moment of inertia. If the star contains a fluid
component that is not coupled to the external torque, as discussed
below, $I$ is less than the star's total moment of inertia.  A
persistent shift in the spin-down rate could arise through an increase
in the external torque (\cite{GIRP}; \cite{DP}; \cite{LEB}) or from a
change in $I$ (\cite{starquakes}; \cite{capacitors}). The external
torque could vary through either a change in the angle $\alpha$
between the spin and magnetic axes, or in the magnitude of the surface
magnetic field, $B$. Gradual growth of the surface field has been
considered (see, \eg, \cite{field_growth}; \cite{MP}), however, no
mechanism has been suggested by which the surface field could suddenly
increase as a consequence of a glitch.  Ruderman (1976) has suggested
that stresses exerted on the inner crust by the superfluid crack the
crust and move plates of material toward the equator. Because the
magnetic field is frozen to the highly-conductive crust, such plate
tectonics would increase $\alpha$. Plate tectonic activity could also
redistribute matter in such a way as to change the orientation of the
principle axes of the star. As the star relaxes to a new 
rotational state, $\alpha$ can change, though the direction of
change is not obvious. 

We now discuss whether the Crab pulsar's behavior can be understood in
terms of sudden changes in $\alpha$. For illustration we use the
vacuum dipole model, in which the rotational evolution is given by
\begin{equation}
\dot{\Omega} = -{B^2 R^6 \sin^2\alpha\over 6 c^3 I} \Omega^3. 
\label{dipolelaw}
\end{equation}
If the alignment angle $\alpha$ changes by $\Delta\alpha$, the
spin-down rate of the star shifts by an amount 
\begin{equation}
{\Delta\dot{\Omega}\over\dot{\Omega}} = {2\Delta\alpha\over\tan\alpha}.
\label{alphashift}
\end{equation}
The average rate of change of $\Delta\dot{\Omega}/\dot{\Omega}$ due to
glitches over 23 years of observations is (\cite{LPS}) 
\begin{equation}
{d\over dt}\langle{\Delta\dot{\Omega}\over\dot{\Omega}}\rangle
          \simeq 3\times 10^{-5}\ {\rm yr}^{-1}. 
\label{crabrate}
\end{equation}
To account for these persistent shifts,
$\alpha$ must increase at an average rate of
\begin{equation}
{\langle\dot{\alpha}\rangle\over\tan\alpha} \equiv {1\over\tau_\alpha} = 
   1.5\times 10^{-5} {\rm rad}\ {\rm yr}^{-1}. 
\label{crabshifts}
\end{equation}
The growth time $\tau_\alpha$ is thus $7\times 10^4$ yr. At this rate of
change, the entire angular shift over the entire $10^3$ yr lifetime of the
star would be only $\sim t_{\rm age}/\tau_\alpha\sim 1/70$ rad,
considerably less than the upper bound of $\pi/2$.  

Another way for the star's spin-down rate to increase is through a
decrease in the effective moment of inertia $I$ on which the external
torque acts. Such a change could occur through either a structural
change of the star, \eg, the star become less oblate, or through a
decoupling of a portion of the star's liquid interior from the
external torque. The change in the spin-down rate, from
eq. [\ref{cons}] would be 
\begin{equation}
{\Delta\dot{\Omega}\over\dot{\Omega}} = -{\Delta I\over I}.
\label{Ishift} 
\end{equation} 
A decrease in $I$ solely through structural readjustment cannot
produce the behavior seen in the Crab pulsar. In this case, angular
momentum conservation would require the star to always spin {\em more}
rapidly than had the glitch not occurred. The large persistent shifts
following the 1975 and 1989 glitches, however, eventually caused the
star to spin {\em less} rapidly than had the glitch not occurred (see
Fig. 1). In principle, the Crab's spin-down rate offsets could be due to
decoupling of a portion of the star's superfluid interior from the
external torque (\cite{capacitors}; \cite{more_capacitors}), though
quantitative agreement of this model with the data has yet to be
demonstrated. 

\section{The Vela Pulsar}

The high levels of glitch activity and timing noise typically
exhibited in older pulsars have prevented the determination of their
braking indices. Recently, however, Lyne \etal\ (1996) attempted to
separate the glitch activity of the Vela pulsar from its underlying
spin-down to obtain the average $\ddot{\Omega}$. The inferred
braking index is $1.4\pm 0.2$, significantly smaller than the braking
indices measured in the younger Crab, PSR B1509-58, and PSR B0540-69,
suggesting evolution of $n_{\rm obs}$. Unlike the determinations of
the braking indices in these young pulsars, which utilized data
intervals that were free of glitches, the measurement of Vela's
$n_{\rm obs}$ employed data containing many glitches. Hence, the value
of Vela's $n_{\rm obs}$ includes the effects of any shifts in
$\dot{\Omega}$ that might arise from glitches.  Here we note that the
small braking index for Vela reported by Lyne \etal\ (1996) could be
due to increases in $\alpha$ of the sort suggested by the Crab's
persistent shifts in spin-down rate.

Suppose that a pulsar spins down according to a braking law of the
form 
\begin{equation}
\dot{\Omega} = -K(t) \Omega^n,
\label{generallaw}
\end{equation}
where, for the vacuum dipole model, $K=B^2 R^6 \sin^2\alpha/6 c^3 I$
and $n=3$. If $K$ depends on time, the observed value of the braking
index, $n_{\rm obs}\equiv\Omega\ddot{\Omega}/\dot{\Omega}^2$, will
differ from $n$ appearing in eq. [\ref{generallaw}].  If $K$ is
changing at an average rate $\dot{K}$, 
differentiating eq. [\ref{generallaw}] in time gives (see, \eg, 
\cite{michels_book}) 
\begin{equation}
n - n_{\rm obs} = -{\Omega\dot{K}\over\dot{\Omega} K} = 
2 t_{sd} {\dot{K}\over K}\ ,
\end{equation}
where $t_{sd}\equiv \Omega/2|\dot{\Omega}|$, the conventional 
spin-down age, differs from the true age if $K$ is evolving. 
Hence, if $K$ is increasing with time, $n_{\rm obs}$ 
will be less than $n$. In the context of the
vacuum dipole model, the growth rate of $\alpha$ required to explain
the braking index of Vela is 
\begin{equation}
{\langle\dot{\alpha}\rangle\over\tan\alpha} \equiv {1\over\tau_\alpha} = 
{3 - n_{\rm obs}\over 4 t_{sd}} =
3.6\times 10^{-5}{\rm rad}\ {\rm yr}^{-1}, 
\label{vela_alpha_rate}
\end{equation}
giving a growth time $\tau_\alpha=3\times 10^4$ yr.  This rate is
remarkably close to the average rate of orientation shifts inferred
from the persistent spin-rate shifts observed in the Crab (see
eq. \ref{crabshifts}). The growth time is consistent with Vela's age,
as $\tau_\alpha<t_{\rm age}$.

The changes in $\alpha$ at the average rate given in eq.
[\ref{vela_alpha_rate}] may occur suddenly, coincident with glitches,
as we are suggesting for the Crab.  Since the average glitch interval
in Vela is $\sim 3$ yr, each glitch would then produce a persistent
offset in the spin-down rate of $\Delta\dot{\Omega}/\dot{\Omega}\sim
2\times 10^{-4}$, comparable to the shifts seen in the Crab. These
persistent shifts would be very difficult to see in Vela; long after a
glitch, $\dot{\Omega}$ typically differs from the average spin-down
rate by $\sim 1$\%, much larger than the expected persistent shifts
assuming the changes in $\alpha$ occur at glitches.

In principle, Vela's lower braking index could be due to a decreasing
moment of inertia (\cite{vela_index}). In this case, 
$\dot{K}/K=-\dot{I}/I$ in the vacuum dipole model, and the required
rate of change is 
\begin{equation}
-{\dot{I}\over I} = {3 - n_{\rm obs} \over 2 t_{sd}}
\simeq {0.8\over t_{sd}}. 
\end{equation}
If $n_{\rm obs}$ remains low over much of the star's life ($\gap
t_{sd}$), then the relative change in the moment of inertia would be
of order unity (see also \cite{michels_book}).  Explanations based on
either decoupling of the interior liquid or changes in the star's
structure cannot accommodate such drastic decreases in $I$. In the
standard picture of the neutron star interior, the only liquid
component that can be decoupled is the inner crust neutron superfluid,
which comprises at most 10\% of the star's total moment of
inertia. Furthermore, structural changes in $I$ are limited by the
deviation from sphericity due to rotation, which is $\Delta I/I\lap
10^{-4}$ for Vela.

\section{Discussion}

We have shown that the persistent offsets in the Crab's spin-down rate
following glitches could be due to sudden glitch-induced increases in
the angle $\alpha$ between the rotational and magnetic axes.
Moreover, a similar average growth rate of $\alpha$ accounts for
the very low braking index of the Vela pulsar. In the vacuum dipole
model, the characteristic growth times of $\alpha$ in these two
pulsars are $\tau_\alpha\sim 7\times 10^4$ yr for the Crab and
$\tau_\alpha\sim 3\times 10^4$ yr for Vela. 

Lyne and Manchester (1988) found that pulsars older than $\sim 10^6$
yr tend to have smaller $\alpha$ than younger pulsars. If our
suggestion that $\alpha$ grows in young pulsars is correct, and the
trend found by Lyne and Manchester (1988) is statistically
significant, then the evolution of $\alpha$ is not monotonic. That is,
in young pulsars ($t_{\rm age}\lap 10^4$ yr) the orientation angle
$\alpha$ may grow, whereas in older pulsars it becomes smaller. 

It is possible that changes in $\alpha$ are responsible for $n_{\rm
obs}<3$ for the Crab, PSR B1509-58, PSR B0540-69. Since the data
intervals used to determine these braking indices contained no
glitches, to explain these values of $n_{\rm obs}$ the alignment angle
$\alpha$ would have to grow {\em between} glitches.

The changes in $\alpha$ inferred from the Crab's persistent offsets
might produce measurable changes in the pulse profile. For example, a
change of $\alpha$ at a glitch could change the duration of the
line-of-sight's traverse through the pulse emission cone. The
magnitude of the associated change in the total pulsed flux is
$\sim|\Delta\alpha|/w_{1/2}$, where $w_{1/2}$ is the half-width of the
emission cone.  Taking $\alpha=86^\circ$, determined from the radio
data (\cite{crab_alpha}), $\Delta\alpha$ for the 1989 glitch (which
exhibited the largest observed offset) is $\simeq 3\times 10^{-3}$ rad
(see eq. \ref{alphashift}). From these numbers we estimate the
relative change in pulsed flux to be $\sim 1$\%. We estimate a
somewhat smaller change, $\sim 0.4$\%, using the value
$\alpha=80^\circ$ obtained by Yadigaroglu and Romani (1996) in their
analysis of gamma ray and radio data. Such flux changes may be
measurable in either x-ray or optical bands.  Sekimoto \etal\ (1995),
using GINGA, found an upper limit on the change in the pulsed x-ray
emission (1-6 keV) across the 1989 glitch of $<1.6$\%. Moreover,
Jones, Smith and Nelson (1980), found that the pulsed optical emission
varied by $\sim 1$\% over 7 yr.

\acknowledgements

We thank G. Baym for helpful comments on this manuscript, and 
A. G. Lyne for providing us with data from the 1989 Crab
glitch.  This work was performed under the auspices of the U.S.
Department of Energy, and was supported in part by NASA EPSCoR Grant
\#291471, and by IGPP at LANL.

\newpage

\figcaption{The behavior of the Crab pulsar surrounding the 1989
glitch at $t_g\simeq 47747$ MJD. The points show the difference
$\delta\Omega(t)\equiv \Omega(t)-\Omega_0(t)$
between the observed spin rate and that expected from a
model to the preglitch behavior $\Omega_0(t)$. A shift in the
spin-down rate of $\Delta\dot{\Omega}/\dot{\Omega}\simeq 4\times
10^{-4}$ persisted until the next glitch. By 40d after the glitch, the
pulsar was spinning slower than had the glitch not occurred.}

\begin{deluxetable}{lcc}
\tablewidth{30pc}
\tablecaption{Glitches of the Crab Pulsar\tablenotemark{a}}
\tablehead{
\colhead{Date} & \colhead{Glitch size} & \colhead{Persistent spin-down
shift} \nl
\colhead{} & \colhead{$10^8\Delta\Omega/\Omega$} &
\colhead{$10^4\Delta\dot{\Omega}/\dot{\Omega}$}}
\startdata
1969 Sep & 0.40 & 0.04 \nl 
1975 Feb & 4.4  & 2    \nl
1981     & -    & 0.1  \nl
1986 Aug & 0.41 & 0.2  \nl
1989 Aug & 8.5  & 4    \nl
1992 Nov\tablenotemark{b}  & -    & -    \nl
\enddata
\tablenotetext{a}{From Lyne, Prichard, \&\ Smith (1993).}
\tablenotetext{b}{The behavior following the 1992 glitch is not yet 
published.}
\end{deluxetable}

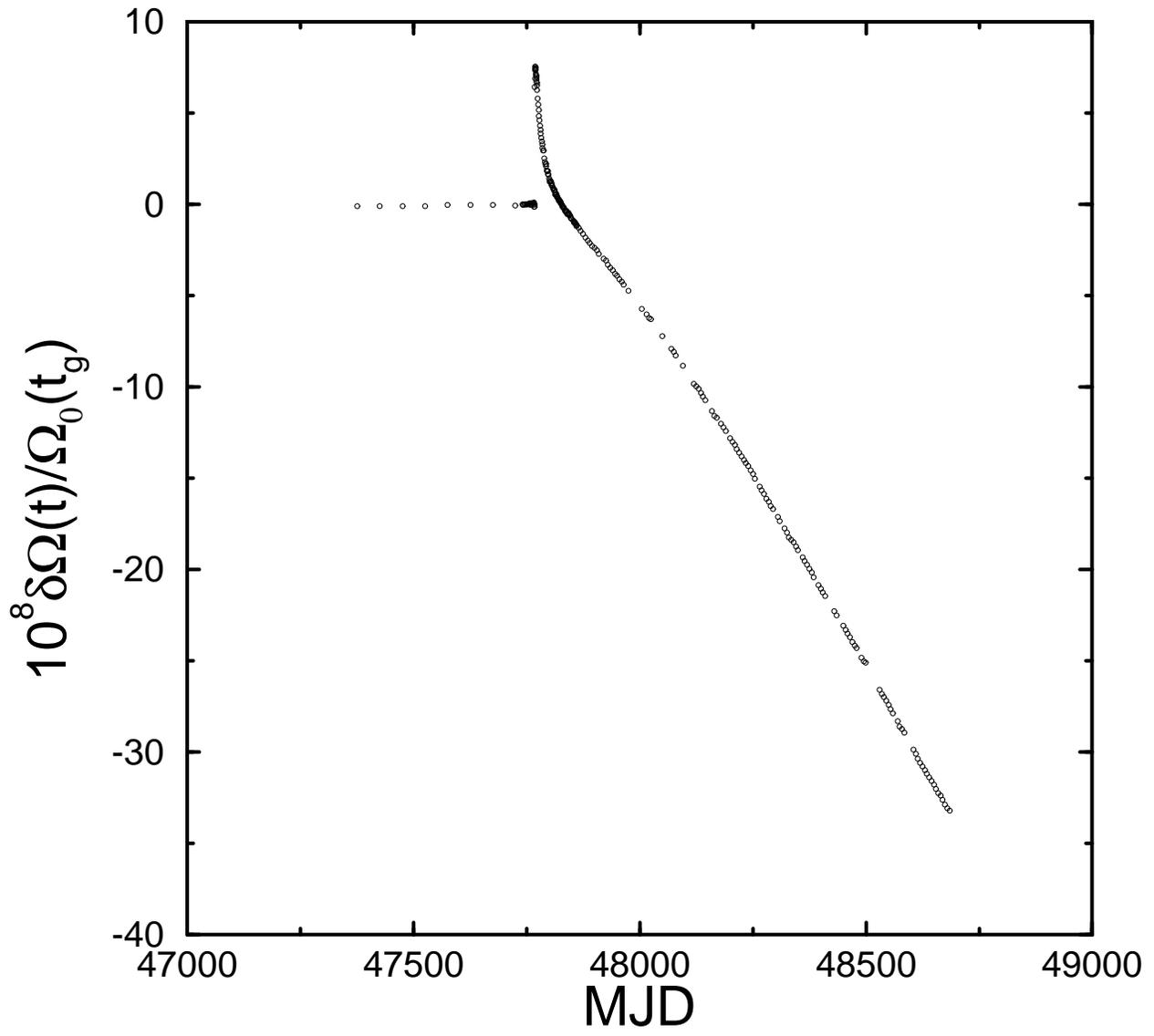
\end{document}